\documentclass[letterpaper,english,aps,prb,amsmath,amssymb,reprint,author-year,author-numerical,floatfix,raggedfooter,superscriptaddress]{revtex4-1}
\usepackage{graphicx}
\graphicspath{{./images/}}
\usepackage{dcolumn}
\usepackage{bm}
\usepackage{mathtools}
\usepackage[caption=false]{subfig}
\usepackage{amssymb}
\expandafter\let\csname equation*\endcsname\relax
\expandafter\let\csname endequation*\endcsname\relax
\usepackage{amsmath}
\usepackage{amsbsy}
\usepackage{booktabs}

\begin{document}

\preprint{AIP/123-QED}

\title{Superconductivity induced by flexural modes in non $\sigma_{\rm h}$-symmetric Dirac-like two-dimensional materials:
       A theoretical study for silicene and germanene}
\author{Massimo V. Fischetti}
\email{max.fischetti@utdallas.edu.}
\affiliation{Department of Materials Science and Engineering, The University of Texas at Dallas\\
             800 W. Campbell Rd., Richardson, TX 75080}
\author{Arup Polley}
\affiliation{Kilby Research Labs, Texas Instruments Inc., 135600 N. Central Expwy., Dallas, TX 75243}

\date{\today}
\begin{abstract}
In two-dimensional crystals that lack symmetry under reflections on the horizontal plane of the lattice 
(non-$\sigma_{\rm h}$-symmetric), electrons can couple to flexural modes (ZA phonons) at
first order. We show that in materials of this type that also exhibit a Dirac-like electron dispersion, the strong coupling can result
in electron pairing mediated by these phonons, as long as the flexural modes are not damped or suppressed by additional interactions with a supporting 
substrate or gate insulator. We consider several models: The weak-coupling limit, which is applicable only
in the case of gapped and parabolic materials, like stanene and HfSe$_{2}$, thanks to the weak coupling; the full gap-equation, solved using the 
constant-gap approximation and considering statically screened interactions; its extensions to energy-dependent gap and to dynamic screening. We argue that in   
the case of silicene and germanene superconductivity mediated by this process can exhibit a critical temperature of a few degrees K, or even a few
tens of degrees K when accounting for the effect of a high-dielectric-constant environment. 
We conclude that the electron/flexural-modes coupling should be included in studies of possible superconductivity in 
non-$\sigma_{\rm h}$-symmetric two-dimensional crystals, even if alternative forms of coupling are considered. 
\end{abstract}


\maketitle

\section{Introduction}
Flexural modes (or "ZA phonons") have always been of great interest in the context of two-dimensional (2D) materials. Their role in undermining the thermodynamic stability of 2D crystals is the basis of the Mermin-Wagner theorem\cite{Peierls_1935,Landau_1937,Mermin_1966,Hohenberg_1967,Coleman_1973}. Their effect on electronic transport has been investigated both in the case of 2D crystals that are symmetric under reflections on the horizontal plane of the lattice ($\sigma_{\rm h}$-symmetric crystals) that permit only their weak coupling to electrons at 
second-order (such as graphene\cite{Mariani_2008,Mariani_2010,Gornyi_2012,Gornyi_2015,Gunst_2016}), as well as in `buckled' 
non-$\sigma_{\rm h}$-symmetric crystals in which a stronger coupling is allowed at first order (such as silicene or germanene, 
for example\cite{Gunst_2016,Fischetti_2016}). The reason of such interest stems from the parabolic dispersion of the ZA phonons, a
peculiarity that results in divergent equilibrium occupation numbers and electron-phonon matrix elements. Recent work has shown how this parabolic 
dispersion is actually renormalized by their anharmonic coupling with in-plane acoustic modes, resulting in a frequency-wavevector relation 
$\omega^{\rm (ZA)}(q)$ proportional to $q^{\eta}$ (where $q$ is the magnitude of the phonon wavevector), with 
a `renormalized' exponent $\eta$ approximately equal to 3/2. This is sufficient to guarantee the thermodynamic stability of the crystals and to weaken significantly the second-order (two-phonon) coupling to electrons in -($\sigma_{\rm h}$-symmetric crystals. 
On the other hand, one of us (MVF) has argued that in non-$\sigma_{\rm h}$-symmetric 2D crystals that exhibit a Dirac-like electron dispersion
the 'renormalized' coupling is still strong enough to affect severely the carrier mobility\cite{Fischetti_2016}.

If, on the one hand, a strong electron/ZA-phonon coupling is unwelcome from an electron-transport perspective, on the other hand such a strong
coupling may suggest the possibility of electron pairing and the emergence of superconductivity. The purpose of this paper is to show that, indeed,
flexural modes may lead to the formation of Cooper pairs in Dirac-like, non-$\sigma_{\rm h}$-symmetric crystals. Obviously, we assume that the flexural
modes are not damped or suppressed by interactions with a supporting substrate or a gate insulator, as discussed in Ref.~\onlinecite{Fischetti_2016}.
Therefore, our discussion applies to free-standing monolayers or layers interacting only weakly (such as via Van der Waals interactions) with the environment.  

We organize our discussion as follows: Keeping in mind the cases of silicene and germanene as significant examples, in Sec.~\ref{sec:non-BCS} we consider
the consequences of the significantly different wavelength dependence of the phonon frequency and of the electron-phonon interactions considered by the
`conventional' Bardeen-Cooper-Schrieffer (BCS) theory\cite{BCS_1957,Pines_1958} and those of interest here. We emphasize the role played by  
Migdal's theorem\cite{Migdal_1958}, the failure of the weak-coupling limit\cite{Kittel,Fetter-Walecka} in our case, and the need to consider numerically 
the full gap-equation, even going beyond McMillan's empirical `strong-coupling' formula\cite{McMillan_1965,McMillan_1968} towards to Eliashberg's
formulation of the problem\cite{Eliashberg_1960,Eliashberg_1961,Ummarino_2013}. 
In Sec.~\ref{sec:silicene}, we briefly review the experimental\cite{Chen_2013} and 
theoretical status regarding the emergence of superconductivity in silicene (and also germanene\cite{Ezawa_2012,Baskaran_2016}) 
-- both phonon-mediated\cite{Wan_2013,Durajski_2015} and non-phonon-mediated\cite{Liu_2013,Zhang_2015}. We then present our results
using the constant-gap approximation, assuming static screening. We later extend them to the solution of the energy-dependent gap also in the case
of dynamic screening and in the presence of monolayers embedded in a dielectric. Our conclusions are presented in Sec.~\ref{sec:Conclusions} and can be summarized by saying that, while not excluding alternative mechanisms that may lead to an efficient electron 
pairing\cite{Takimoto_2004,Loktev_2004,Yada_2005,Kubo_2007,Kuroki_2008,Graser_2009}, flexural modes should be considered in any study that deals
with superconductivity in Dirac-like, non-$\sigma_{\rm h}$-symmetric 2D crystals.  
\vspace*{-0.00cm}
\section{The electron/ZA-phonon coupling}
\label{sec:non-BCS}
\vspace*{-0.00cm}
\subsection{Electron-phonon interaction}
\label{subsec:elphon}
The potential energy associated with the phonon-mediated electron-electron interaction we consider here has the form:
\begin{equation}
V^{\rm (ep)}({\bf q}) \ = \frac{\hbar \omega({\bf q}) M({\bf q})^{2}}
                                 {[E({\bf k})-E({\bf k}+{\bf q})]^{2} - \hbar^{2} \omega({\bf q})^{2}} \ ,
\label{eq:Vep_1}
\end{equation}
where ${\bf k}$ is the two-dimensional electron wavevector, $E({\bf k})$ is the electron dispersion, and $\omega({\bf q})$ is the frequency of 
ZA phonons of wavevector ${\bf q}$. (We omit for simplicity the superscript `ZA')
In the case of non-symmetric 2D materials, the electron-phonon term originates from the first-order coupling of electrons
with out-of-plane acoustic phonons and has the form:
\begin{equation}
M({\bf q})^{2} \ = \ \frac { \hbar [DK({\bf k},{\bf k}+{\bf q})]^{2} } 
                          {2 \rho \ \omega({\bf q}) \ [\epsilon({\bf q},\omega)/\epsilon_{\rm s}]^{2} }  \ ,
\label{eq:Eep_1}
\end{equation}
where $\rho$ is the (2D) mass density of the crystal.  
The interaction is assumed to be statically or dynamically screened, thanks to the factor $[\epsilon_{\rm s}/\epsilon(q,\omega)]^{2}$,
where $\epsilon(q,\omega)$ and $\epsilon_{\rm s}$ are the dielectric function and static dielectric constant of the crystal, respectively.
The `deformation potential' $DK({\bf k},{\bf k}+{\bf q})$, is proportional to $\Delta_{\rm ZA} q$ for `gapped' non-symmetric 2D materials 
($\Delta_{\rm ZA}$ being usually called the `acoustic deformation potential'), whereas it is independent of energy in 
Dirac-like 2D materials, although with an importance dependence, $\sin(\phi/2)$, on the scattering angle $\phi$, as
shown in Ref.~\onlinecite{Fischetti_2016}.
At the low temperatures of interest here, the Bose-Einstein phonon-occupation term $N({\bf q})$ can be ignored in Eq.~(\ref{eq:Vep_1}).\\

\subsection{BCS-like and non-BCS-like interaction}
\label{subsec:nonBCS}
Despite its `familiar' form, Eq.~(\ref{eq:Vep_1}) hides significant differences with respect to the `conventional' BCS theory. 
In this latter case, the matrix element
$M({\bf q})$ grows with increasing magnitude, $q$, of the wavevector of acoustic (or Debye) phonons, so that large-energy phonons control the coupling.
Indeed, since their frequency grows as $c_{\rm s}q$, the process is mainly controlled by 
zone-edge modes of the Debye frequency $\omega_{\rm D} = c_{\rm s} q_{\rm BZ}$.
(Here $c_{\rm s}$ is some angle-averaged acoustic velocity and $q_{\rm BZ} = 2\pi/a_{0}$ the equally-angle-averaged wavevector at the edge
of the Brillouin zone expressed in terms of the lattice constant $a_{0}$.) 
This implies the existence of a relative large region of ${\bf k}$-space in which $|E({\bf k})-E({\bf k}+{\bf q})| \ll \hbar \omega({\bf q})$
and Eq.~(\ref{eq:Vep_1}) represents an attractive interaction. Moreover,
Migdal's theorem\cite{Migdal_1958} guarantees that the singularities in Eq.~(\ref{eq:Vep_1}) ({\em i.e.}, the poles 
$E({\bf k})-E({\bf k}+{\bf q}) = \pm \hbar \omega({\bf q})$) give a negligible contribution, 
since not only $\upsilon_{\rm F} \gg c_{\rm s}$ in metals but, also,
they occur in the small-$q$ region in which the interaction $\sim |M({\bf q})|^{2}$ is weak. The net result is that when considering the effective
phonon-mediated electron-electron interaction, given by Eq.~(\ref{eq:Vep_1}), the effect of repulsive terms and poles can be neglected and we can consider
only the attractive part of the interaction:  
\begin{equation}
V^{\rm (ep)}({\bf q}) \ \sim \ - \ \frac{M({\bf q})^{2}}{\hbar \omega({\bf q})} \ .
\label{eq:Vep_BCS}
\end{equation}

When considering the electron/ZA-phonon interaction in gapped, parabolic, non-$\sigma_{\rm h}$-symmetric 2D materials, the situation is very similar:
Although the phonon frequency vanishes faster than $q$ as $q \rightarrow 0$, the strength of the interaction grows with increasing $q$. Therefore,
we expect a BCS-like behavior, although with a weak coupling. However, the picture is completely different when considering Dirac-like materials. In this case, 
the matrix element $M({\bf q})$ grows with {\em decreasing} $q$, actually diverging in the limit $q \rightarrow 0$ for interactions left unscreened.
Therefore, low-energy ZA-phonons control the coupling. Moreover, the singularities are not even `poles' (since we are assuming $\omega({\bf q}) = b q^{3/2}$) 
and they may give non-negligible contributions, at least in principle, so that the validity of Migdal's theorem is not guaranteed. Of course,
all this also implies that whether or not the interaction given by Eq.~(\ref{eq:Vep_1}) can ever be attractive must be established with 
careful calculations. In any event, we expect a strong interaction at low densities, since the Fermi surface/line probes the region of a 
strong interaction near the phonon $\Gamma$ symmetry-point (and, so, small Fermi wavevectors). 
This is the opposite behavior of what is seen in the conventional BCS case.

Another significant difference between our situation and those handled by the BCS theory lies in the role played by dielectric screening. 
In the BCS case, usually applied to metals, the large plasma frequency, much larger than the Debye frequency and the superconducting gap, 
justifies the use of static screening. In our systems, considering 
statically screened interactions would also be appropriate in the normal state: Thanks to the very low frequency of the ZA phonons of interest,
the plasma frequency, even if wavevector dependent and vanishing at long wavelength (the case of 2D systems), would also be large enough to justify the full response of the electron gas. However, in the superconducting state electrons exchange an additional energy of the order of the superconducting gap
at the Fermi energy, $\Delta(k_{\rm F})$. The associated frequency, $\Delta(k_{\rm F})/\hbar$, even if small in the weak-coupling regime,
could overcome the plasma frequency, thus rendering dynamic effects extremely important. Therefore, we shall consider static screening at first, but we
shall consider dynamically screened interactions in Sec.~\ref{subsec:dynamic}.  

This discussion shows that the problem we are facing is far from being trivial. Indeed, we must face problems similar to those already discussed in
the literature: The validity of Migdal's theorem in Dirac-like materials has been proven in the case of graphene by Roy and
coworkers\cite{Roy_2014}, but only for `conventional' interactions whose strength grows with decreasing wavelength. In our case, it is true that
the so-called `adiabatic' limit is reached also at small carrier densities, since $\upsilon_{\rm F} \gg c_{\rm s}$. However, the role of the singularities
in Eq.~(\ref{eq:Vep_1}) remains unclear. (We note that Migdal's theorem
has been claimed to hold even in the anti-adiabatic regime in some cases\cite{Takada_1992})
Interactions that grow as $1/q$ at long wavelength and that may violate Migdal's theorem  
have also been considered extensively in the context of high-$T_{\rm c}$ 
superconductors\cite{Kulic_1994,Zeyher_1996,Pietronero_1995,Danylenko_2000,Palistrant_2006}. Finally, 
superconducting states with symmetries significantly different from the BCS $s$-wave pairing, as reviewed by
Tsuei and Kirtley\cite{Tsuei_2000}, may dominate the picture in our case. Indeed, as we shall see below, such mechanisms have been
proposed in the case of silicene. Therefore, we shall now proceed in steps, considering increasingly complicated models to gain some 
insight on what role the flexural modes may play.

\subsection{Approximations and physical models}
\label{sebsec:Approx}
Before proceeding, in this brief subsection we summarize physical models and approximations we have embraced throughout the paper.

First, even if/when attractive, the phonon-mediated electron-electron interaction must overcome the Coulomb repulsion whose potential energy is, 
in principle, given by
\begin{equation}
V^{\rm (C)}({\bf q}) \ = \  \frac{e^{2}} { 2 \epsilon({\bf q}, \omega) q } \ .
\label{eq:Coulomb}
\end{equation}
This interaction presents a divergence as $q \rightarrow 0$, easily circumvented by accounting for dielectric screening, and as $q \rightarrow \infty$.
This latter divergence has been shown to be removed by a many-body renormalization of the Fermi velocity\cite{Roy_2014} in Dirac-like materials. More generally, it is removed either by employing a high-$q$ cutoff (here we shall use the zone-edge $q_{\rm BZ}$) or the use of the the Morel-Anderson 
pseudopotential\cite{Morel_1962}. In any event, we shall show below that the superconducting gap, $\Delta({\bf k})$, vanishes
at large energies, so that this issue is not critical.   

Regarding the electronic band structure and phonon dispersion, we shall adopt model-dispersions assumed to be valid throughout the Brillouin zone. 
Therefore, for
gapped, parabolic materials, we shall assume an isotropic dispersion $E({\bf k)} = \hbar^{2}k^{2}/(2 m^{\ast})$ with an effective mass $m^{\ast}$, whereas
for Dirac-like 2D crystals we shall assume $E({\bf k}) = \hbar \upsilon_{\rm F}k$. Since we are interested in a region of the Brillouin zone
in which $E({\bf k})$ is close to the Fermi energy, $E_{\rm F}$, these expressions are valid as long as the Fermi energy (and so the carrier density) is
low enough to be in the appropriate parabolic or linear region and, for Dirac-like materials, large enough to be left unaffected by the opening
of a gap due to the spin-orbit interaction. In silicene, this interaction results in the opening of a very small gap of about 1.5~meV in the buckled 
structure\cite{{Liu_2011},{Liu_2011a},{Yakovin_2017}} at the Dirac point. This should not affect significantly our results for electron densities larger
than about $6 \times 10^{8}$ cm$^{-2}$, in practice equivalent to an undoped/ungated case. 
On the contrary, for germanene, the small gap of about 2~meV predicted to occur in the planar (un-buckled) structure\cite{Singh_2017} grows to 
about 24~meV in the buckled structure\cite{{Liu_2011},{Liu_2011a}} and this would certainly modify the Fermi surface for densities smaller than
about $1.5 \times 10^{11}$ cm$^{-2}$. Yet, as we shall see below, the largest values for the superconductivity gap and transition temperature will be
found at densities that are within the range of validity of our model, around the high-$10^{12}$/low-$10^{13}$ cm$^{-2}$. 
These values are large enough to be left unaffected by the spin-orbit interaction, yet small enough to be satisfactorily described by a pure Dirac-like electron dispersion. 

Similarly, we shall assume a ZA-phonon dispersion of the form $bq^{3/2}$ -- the parameter $b$ being fixed by the
condition $bq_{\rm BZ}^{2/3} = \omega_{\rm D}$ ($\approx$5~meV for SnI and HfSe$_{2}$, 15~meV for silicene, and 9~meV for germanene)
throughout the entire Brillouin zone. 
This latter assumption should not affect the results in any significant way, since most of the `action' happens at low $q$. However, we should note that assuming a pure Dirac-like dispersion (and also a constant
electron/ZA-phonon deformation potential) may depress the values of the calculated superconductivity gap, since the repulsive part of the phonon-mediated effective electron-electron interaction will not vanish as fast as when assuming a pure Dirac dispersion. Yet, even in this case, for densities in the range that is realistically obtained by gating or doping, our assumptions should be approximately satisfactory.     

Regarding dielectric screening of both the electron-phonon and the Coulomb interaction, we have used
Wunsh'\cite{Wunsch_2006} or Stern's\cite{Stern_1967} expressions for $\epsilon(q,\omega)$ for Dirac-like and gapped (parabolic) materials,
respectively. Screening of the nonpolar interaction between electrons and acoustic phonons has been discussed at length in the past, with Cardona and 
Christensen showing the necessity of the such a screening for the dilatation (hydrostatic) deformation potential\cite{Cardona_1987}. 
However, Tanatar has treated all nonpolar interactions as screened 1D structures\cite{Tanatar_1993} and, more recently,   
such a form for dielectric screening of the electron/acoustic-phonon interaction has been considered also in 2D materials by 
Kaasbjerg {\em et al.}\cite{Kaasbjerg_2013} for the {\em Normal} (as opposite to {\em Umklapp}) processes we are considering here. A `parochial' but 
exhaustive discussion of the history behind this `screening problem' is given in Ref.~\onlinecite{Fischetti_1993}, paper that also provides
several additional references. As mentioned above, we shall consider statically screened interactions at first, but we shall extend our study
to include dynamic-screening effects in Sec.~\ref{subsec:dynamic}.

As mentioned before, we approximate the deformation potential $D({\bf k},{\bf k}')$ appearing in Eq.~(\ref{eq:Vep_1}) with
$\Delta_{\rm ZA} q$ (with $q=|{\bf k}-{\bf k}'|$) for parabolic, gapped materials; for Dirac-like crystals, instead, 
$D({\bf k},{\bf k}') \approx DK_{0} \sin (\phi/2)$. This is a satisfactory approximation when spin-orbit interaction is ignored. However, we
expect it to remain a good approximation whenever the electron energy can be approximated by a linear Dirac-like dispersion, as discussed above
in the context of the band structure. The different forms taken by $D({\bf k},{\bf k}')$ in these two different types of materials has
been discussed in Ref.~\onlinecite{Fischetti_2016} in the context of electron transport. We shall see below that similar considerations apply also
in the context of superconductivity.  

Finally, we shall limit ourselves to a simple $s$-wave pairing and all physical and material parameters are identical to those used in 
Ref.\onlinecite{Fischetti_2016}. 

\subsection{The weak-coupling limit}
\label{subsec:Weakcoupling}
As our first step, we consider the `usual' BCS-like weak-coupling limit. This is a realistic approximation for the quasi-conventional case of
parabolic. gapped materials; on the other hand, we shall see that it is completely inadequate in the case of Dirac-like crystals.
 
We start by considering the equation for the superconducting gap. In the infinite-volume normalization, this can be written as:
\begin{multline}
\Delta({\bf k}) \ = \\ 
- \int \frac{\rm{d}{\bf k}'}{(2 \pi)^{2}} \ V({\bf k}-{\bf k}') \ {\mathcal{I}}({\bf k},{\bf k}')
                                            \ \frac{ \Delta({{\bf k}'}) } { 2 W({{\bf k}'}) } 
                                                 \tanh \left ( \frac{W({\bf k}')}{2k_{\rm B}T} \right ) \ ,
\label{eq:gap_BCS}
\end{multline}
where $k_{\rm B}$ is the Boltzmann's constant, $T$ the temperature,
$V(q)$ is the total potential energy given by the sum of the electron-phonon and Coulomb potential energies, 
$V^{\rm (C)}({\bf q}) + V^{\rm (ep)}({\bf q})$, and $W({\bf k}) = \sqrt{|E({\bf k})-E_{\rm F}|^{2} + \Delta({\bf k})^{2}}$ 
is the renormalized energy measured from the Fermi level. The `overlap factor' ${\mathcal{I}}({\bf k},{\bf k}')$ is assumed to be $(1+\cos \phi)/2$, 
where $\phi$ is the scattering angle, for Dirac-like materials, unity otherwise.
Thanks to the isotropy of the electronic dispersion and having assumed $s$-wave pairing, 
all quantities in the expression above depend only on the magnitude $k$ of ${\bf k}$, and in the following the notation will be simplified 
accordingly.\cite{Note2}
In the weak-coupling limit, $\Delta(k_{\rm F}) \ll E_{\rm F}$, and a zero temperature,
the well-known approximate solutions for the gap at the Fermi energy, $\Delta(k_{\rm F})$, is given by:
\begin{equation}
\Delta(k_{\rm F }) \ \approx \ 2 \hbar \omega_{\rm D} \exp \left ( \frac{1}{\nu V_{\rm{eff}}} \right ) \ ,
\label{eq:gap_approx_3}
\end{equation}
where $\nu$ is the density of states {\em per spin-state} at the Fermi energy and
\begin{equation}
V_{\rm{eff}} \ = \ \int_{0}^{\pi} \ \frac{\rm{d}\phi}{\pi} \ {\mathcal{I}}(\phi) \ 
                           V \left [ 2 k_{\rm{F}} \sin (\phi/2) \right ] \ 
\label{eq:Veff}  
\end{equation}
is the potential energy averaged over the Fermi surface/line. Of course, Eq.~(\ref{eq:gap_approx_3}) is valid only when $\nu V_{\rm eff}$ is negative.
From Eq.~(\ref{eq:gap_BCS}), setting the gap to zero one obtains a similar approximate `weak-limit' expression for the transition temperature:
\begin{equation}
k_{\rm B} T_{\rm c} \ \approx \  1.13 \ \hbar \omega_{\rm D} \exp \left ( \frac{1}{\nu V_{\rm eff}} \right ) \ . 
\label{eq:Tc1}
\end{equation}

Whenever the electron density is large enough to render static screening applicable (that is: whenever the 2D plasma
frequency is larger than the phonon energy), we can express the dielectric function as:
\begin{equation}
\epsilon({\bf q},0) \ = \ \epsilon_{\rm s} \left ( 1 + \frac{\beta}{q} \right ) \ ,
\label{eq:eps_static}
\end{equation}  
where 
\begin{equation}
\beta \ = \ \frac{ e^{2}gE_{\rm F}}{4 \pi \hbar^{2} \upsilon_{\rm F}^{2} \epsilon_{\rm s}} \ = \
             \frac{e^{2}}{2 \epsilon_{\rm s}\hbar \upsilon_{\rm F}} \left ( \frac{gn}{\pi} \right )^{1/2} 
\label{eq:eps_static_D}
\end{equation}
for Dirac-like materials ($g$ is the spin and valley degeneracy), and
\begin{equation}
\beta \ = \ \frac{e^{2}m^{\ast}}{2 \pi \epsilon_{\rm s} \hbar^{2}}
\label{eq:eps_static_P}
\end{equation}
for gapped, parabolic materials.

For parabolic materials, let's consider the limit $\beta \gg 2 k_{\rm F}$. Since $\beta \sim 10^{10}$ m$^{-1}$ 
for $m^{\ast} = m_{0}$ and $\epsilon_{\rm s} = 10 \ \epsilon_{0}$,  this condition is met
at densities $n = k_{\rm F}^{2}/(2 \pi) \ll \beta^{2}/(8 \pi) \sim 10^{14}$ cm$^{-2}$. Therefore, taking this limit should be
satisfactory in all reasonable cases\cite{Note1}. In this limit, $\beta/(2 k_{\rm F}) \gg 1$, and
recalling that the density of states (per spin state) at Fermi surface is $m^{\ast}/(2 \pi \hbar^{2})$, we have:
\begin{equation}
\nu V^{\rm (tot)}_{\rm eff} \ \sim \ 
     - \frac{4 \Delta_{\rm ZA}^{2} k_{\rm F} \hbar^{2} \epsilon_{\rm s}^{2}}
                {e^{4} \rho \ b^{2} m^{\ast}} + 1/2  \ . 
\label{eq:Veff_tot_large_a}
\end{equation}

We have considered the two parabolic materials studied in Ref.~\onlinecite{Fischetti_2016}: Iodine-functionalized monolayer tin, SnI, and
a transition metal dichalcogenide with a stable tetragonal structure (T) at room temperature, HfSe$_{2}$. 
With the parameters listed in that reference, 
$T_{\rm c} \approx$ 4~K ($\Delta \approx$ 0.3~meV) at $n = 10^{14}$ cm$^{-2}$, for SnI, 
assuming a static dielectric constant $\epsilon_{\rm s}$=4 $\epsilon_{0}$.

For HfSe$_{2}$, an attractive interaction is obtained 
only at unreasonably high (metallic) densities ($> 10^{15}$ cm$^{-2}$). 
Note that the ZA deformation potential $\Delta_{\rm ZA}$ reported in Ref.~\onlinecite{Fischetti_2016} (and used here) is quite small in both
materials, 1.6~eV and 1.8~eV in SnI and HfSe$_{2}$, respectively. 
\begin{figure}[tb]
\includegraphics[width=8.0cm]{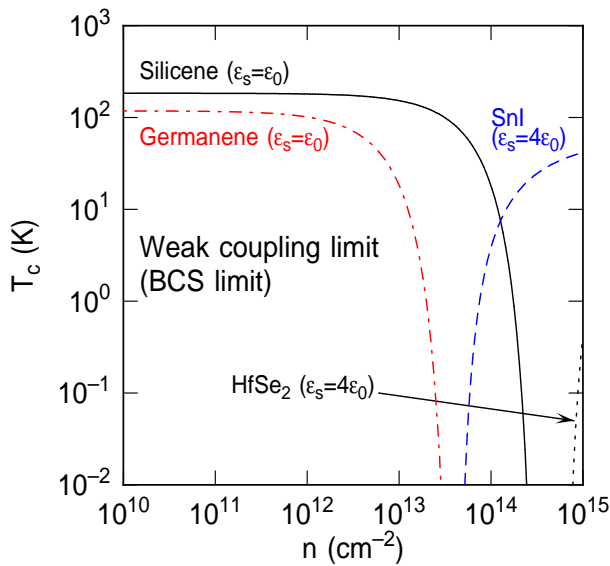}
\caption{Normal-superconducting transition temperature, $T_{\rm c}$, calculated for parabolic, gapped materials (SnI and HfSe$_{2}$) and Dirac-like
silicene and germanene using the weak-coupling limit. Note the usual density dependence and weak effect seen for parabolic materials in which
superconductivity is expected only at unrealistically high metallic carrier densities. On the contrary, for Dirac-like materials, the large values obtained
for $T_{\rm c}$ imply a failure of the weak-coupling approximation.}
\label{fig:Weak-coupling}
\end{figure}

The transition temperature, $T_{\rm c}$, for SnI and HfSe$_{2}$ is shown in Fig.~\ref{fig:Weak-coupling} as a function of the
carrier density $n$. Note that a high electron density is required to boost the attractive effective electron/ZA-phonon interaction, 
as seen in the $k_{\rm F}$ dependence of Eq.~(\ref{eq:Veff_tot_large_a}). 

A final interesting observation in the context of parabolic materials can be made regarding the negligible role played by in-plane modes 
(LA/TA phonons). Indeed, in the same limit $\beta \gg 2k_{\rm F}$, the electron-electron interaction energy due to the coupling with the in-plane modes
takes the form:
\begin{equation}
\nu V^{\rm (ep)}_{\rm eff} \ \sim \ - \frac{m^{\ast} \Delta_{\rm LA}^{2} k_{\rm F}^{2}}{2 \pi \hbar^{2} \rho \ c_{\rm s}^{2} \beta^{2}} \ .
\label{eq:nuVeff_LA_large_a}
\end{equation}
This quantity, rather than being of the order of unity or larger (as in the case of ZA phonons), is of the order of $10^{-2}$-to-$10^{-1}$
at best, even at an unrealistically high `metallic density ($10^{15}$ cm$^{-2}$) and with high deformation potentials (5 eV). At
reasonable -- albeit still very high -- densities, $n \sim 10^{14}$ cm$^{-2}$, one would need $\Delta_{\rm LA} >$ 25 eV to reach unity for
this quantity. 

The main conclusion of this discussion on gapped materials is that it is indeed the strong electron/ZA-phonon coupling 
(due to the superlinear dispersion $\sim q^{3/2}$) that results in superconductivity, albeit at temperatures 
that are not too exciting and at unrealistically high -- almost metallic -- carrier densities. 
Such a relatively weak coupling and the similarity with the conventional BCS theory renders our weak-coupling estimates
reliable for these gapped, non-$\sigma_{\rm h}$-symmetric 2D materials.  

The different form that the electron-phonon matrix element takes in Dirac-like materials results in a dramatically different picture.
Here, we first consider the weak-coupling limit also for these materials. This should be viewed as no more than an exercise
requiring more accurate solutions of the gap equation, since
the `strong coupling' results we shall find hint at a failure of this limit. Nevertheless, this exercise will show the qualitatively correct main trends.
 
In this case, it is customary to assume an in-plane dielectric constant equal to the dielectric constant of the surrounding environment.
For free-standing layers, this assumption implies rather large value of 
$a = \beta/(2k_{\rm F}) = e^{2}/(2\pi \hbar \upsilon_{\rm F} \epsilon_{0}) \approx$ 8.5 for both silicene and germanene. Thus, it is reasonable
to take the large-$a$ limit also in this case. 
Since $\nu = k_{\rm F}/(2 \pi \hbar \upsilon_{\rm F})$ (density of states per spin and valley)
and $\beta = g e^{2} k_{\rm F}/(2 \pi \hbar \upsilon_{\rm F})$ (where $g$=4 is the typical
valley and spin degeneracy of hexagonal Dirac-like 2D crystals), we find for the total interaction energy:
\vspace*{-0.25cm}
\begin{equation}
\nu V^{\rm (tot)}_{\rm eff} \ \sim \  - \frac{(DK_{0})^{2} \hbar \upsilon_{\rm F} \epsilon_{0}^{2} } 
                                          { 12 e^{4} k_{\rm F}^{2} \rho \ b^{2} } \ + \ \frac{1}{16} \ .
\label{eq:nuVeff_tot_D}
\end{equation}
Using the parameters given in Ref.~\onlinecite{Fischetti_2016} for silicene and germanene, we find that
that $\nu V^{\rm (tot)}_{\rm eff}$ is negative for all realistic conditions. Actually, it increases with decreasing carrier density, since in the limit
of zero carrier density the electron-phonon interaction becomes unscreened and diverges as $1/(k_{\rm F}\beta) \sim k_{\rm F}^{-2}$, whereas the  
the Coulomb repulsive term remain constant. With $\hbar \omega_{\rm D}$ = 15~meV for silicene and 9~meV for
germanene, at a density of $10^{12}$ electrons/cm$^{2}$ we obtain $\Delta \approx$ 11~meV and $T_{\rm c} \approx$ 140~K for silicene
and $\Delta \approx$ 0.5~meV and $T_{\rm c} \approx$ 7~K for germanene. 
The transition temperature, $T_{\rm c}$, for silicene and germanene is also shown in Fig.~\ref{fig:Weak-coupling} as a function of the
carrier density $n$. As we have anticipated above, we see a
behavior that is exactly the opposite of what we see in the case of parabolic materials:
As we have remarked above, this is a consequence of the fact that the electron/ZA-phonon matrix element grows at longer wavelengths, 
actually diverging in the unscreened, zero-density limit, $q \sim 2k_{\rm F} \rightarrow 0$. The fact that the transition temperature, 
$T_{\rm c}$, saturates at some maximum value, $\sim \hbar \omega_{\rm D}/k_{\rm B}$, is just an artifact of the weak-coupling expression, Eq.~(\ref{eq:Tc1}). 
This comes from the initial assumption that, if the gap $\Delta$ is small, then the integral of the full gap-equation gives a contribution
only over a shell of thickness $2 \hbar \omega_{\rm D}$ around the Fermi surface. Clearly, if the contributions come from a narrower region of
${\bf k}$-space, the weak-coupling limit misses this fact altogether. Therefore, the estimates above may be viewed as optimistic {\em upper bounds}
for the gap and transition temperature.

\subsection{McMillan's strong-coupling formula}
\label{sebsec:Strongcoupling}
Given the strong coupling and large (unreasonable?) values of $T_{\rm c}$ we find at low densities, it is necessary to consider
the expression one can obtain using Eliashberg's theory\cite{{Eliashberg_1960},{Eliashberg_1961}} in the strong-coupling limit. This theory is
probably the state-of-the-art for `conventional' superconductivity, although the results may differ when considering the long-wavelength
coupling we have to deal with in our cases.

Deferring the task of finding a full solution of the gap equation to the next section, here we consider a `popular' expression used in the 
strong-coupling limit.
\begin{figure}[tb]
\includegraphics[width=8.0cm]{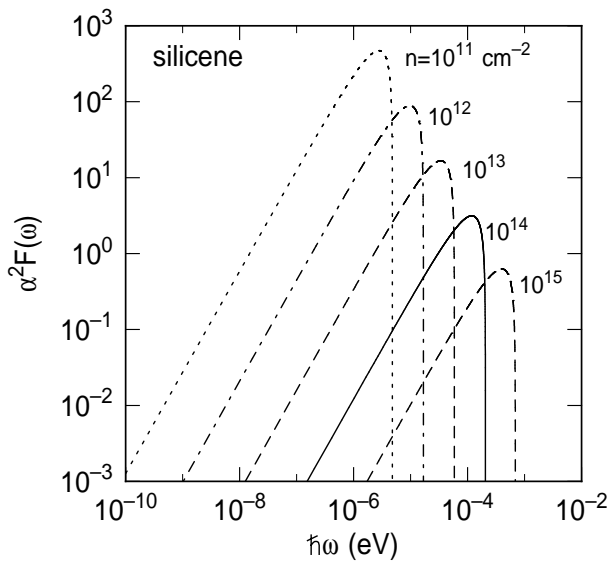}
\caption{Eliashberg's electron-phonon spectral function calculated as a function of frequency at various carrier densities for silicene.}
\label{fig:Elias_aF2}
\end{figure}
\begin{figure}[tb]
\includegraphics[width=8.0cm]{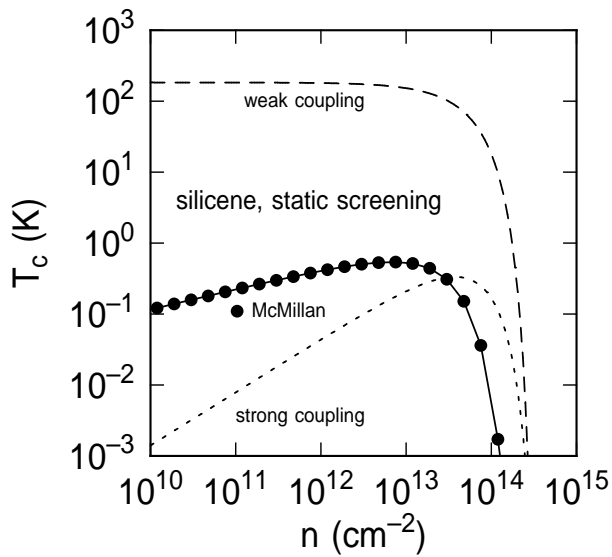}
\caption{Transition temperature calculated for silicene using the weak-coupling limit (dashed lines), the
strong-coupling fit, Eq.~(\ref{eq:Elias_Tc}) of the text (dotted lines), and McMillan's approximation to solve the full gap equation 
within the constant-gap approximation (symbols).}
\label{fig:Strong_coupling}
\end{figure}
In this case, assuming a $k$-independent gap, the quantity of interest is the Eliashberg's electron-phonon spectral 
function\cite{{Eliashberg_1960},{Eliashberg_1961}} (see also the review by Ummarino\cite{Ummarino_2013}):
\begin{equation}
\alpha^{2} F(\omega) = \nu \int_{0}^{\pi} \frac{{\rm d} \phi}{\pi} \ \frac{1+\cos \phi}{2} \
                    \omega(q) \ V^{\rm (ep)}(q) \ \delta [ \omega - \omega(q) ] \ .
\label{eq:aF}
\end{equation}
with $q = 2k_{\rm F} \sin(\phi/2)$. In our case this is:
\begin{equation}
\alpha^{2} F(\omega) = \nu \frac{2}{3} \frac{(DK_{0})^{2}}{\pi \rho \omega_{\rm F}^{2}} 
                         \frac{(\omega/\omega_{\rm F})^{4/3} [1-(\omega/\omega_{\rm F})^{4/3}]}
                              {[(\omega/\omega_{\rm F})^{2/3}+\beta/(2k_{\rm F})]^{2}} \ ,
\label{eq:aF2}
\end{equation}
where $\omega_{\rm F} = b (2k_{\rm F})^{3/2}$. Figure~\ref{fig:Elias_aF2} shows this function at various values of the 
(unrenormalized) density $n$. Note how this function, and so the coupling, is dominated by long-wavelength phonons, especially as $n$ decreases.
 
The strength of the interaction is given by the dimensionless coupling constant:
\begin{equation}
\lambda = 2 \int_{0}^{\omega_{\rm D}} {\rm d} \omega \ \frac{\alpha^{2} F(\omega)}{\omega}  \ ,
\label{eq:lambda}
\end{equation}
which is obviously equal to $-2 \nu V^{\rm (ep)}_{\rm eff}$. 

McMillan\cite{McMillan_1968} has provided an empirical estimate for the transition temperature that, in a slightly revised form given by Allen 
and Dynes\cite{Allen_1975}, takes the form:
\begin{equation}
k_{\rm B} T_{\rm c} = \frac{\hbar \omega_{\rm log}}{1.2} \exp \left [ - \frac{1.04 (1+\lambda)}{\lambda - \mu^{\ast} (1+0.62 \lambda)}
                                      \right ] \ ,
\label{eq:Elias_Tc}
\end{equation}
where $\mu^{\ast}$ is an effective repulsive Coulomb term. As mentioned before, this is often taken to be the Morel-Anderson potential\cite{Morel_1962}. 
However, here we take it as $\nu V^{\rm (C)}_{\rm eff}$. The frequency $\omega_{\rm log}$ is the logarithmic average of the
phonon frequencies involved in the coupling,
\begin{equation}
\omega_{\rm log} = \exp \left [\frac{2}{\lambda} \int_{0}^{\omega_{\rm D}} {\rm d} \log(\omega) \ \frac{\alpha^{2} F(\omega)}{\omega}  \right ] \ .
\label{eq:omega_log}
\end{equation}

The dotted line in Fig.~\ref{fig:Strong_coupling} shows this limit for silicene, compared to the weak-coupling limit. The maximum values of the
transition temperature is reduced. Moreover, the qualitative features are significantly different, showing a reduction of the 
transition temperature at lower carrier densities, as a result of the reduced thickness of the Fermi-shell.

Ummarino\cite{Ummarino_2013} has stressed the fact that the McMillan formula, Eq.~(\ref{eq:Elias_Tc}), is no more than an empirical fit to a limited
set of results that McMillan obtained solving Eliashberg's equations for a few metals\cite{McMillan_1965,McMillan_1968}. Therefore, we must
confirm these results by tackling the harder problem of solving the full gap equation.

\section{Silicene and germanene}
\label{sec:silicene}

Here and in the following we shall consider only non-$\sigma_{\rm h}$-symmetric 2D materials with a Dirac-like electron dispersion. Indeed,
we have argued that the weak-coupling limit describes satisfactorily the case of parabolic, gapped materials. We consider specifically
the interesting cases of silicene and germanene, since the possible emergence of superconductivity in these materials has already been studied (and
even possibly experimentally observed in silicene).

\subsection{Superconductivity in silicene}
\label{subsec:silicene}
Chen {\em et al.}\cite{Chen_2013} have reported the experimental observation of superconductivity in silicene on (111) Ag. 
Superconductivity has also predicted by Wan {\em et al.}\cite{Wan_2013} on the basis 
of phonon-mediated processes in biaxially strained silicene. A similar study has been presented also by Durajski and co-workers\cite{Durajski_2015}, 
also for biaxially-strained silicene. The 
experimentally observed gap, $\Delta$ is about 35~meV, but it disappears at a temperatures of only 35-40~K, 
a gap/temperature ratio that is inconsistent with the universal prediction of the conventional BCS theory. 
Chen and coworkers\cite{Chen_2013} speculate that this is either an artifact due to the fact the temperature of their STM 
tip differs significantly from the sample temperature or, instead and more intriguing, that it indicates that silicene is not an $s$-wave 
superconductors and conventional BCS theory does not apply. However, the experimental situation is quite complex, since measurements were performed 
with STM on silicene supported by (111) Ag and additional effects, such as electric-field-induced interface superconductivity\cite{Uchihashi_2017} 
of Ag may also play a role. Indeed, Zhang and coworkers have speculated about electric-field-induced superconductivity for 
silicene\cite{Zhang_2015} and Zhang and co-workers themselves and Liu {\em et al.}\cite{Liu_2013} have investigated non-$s$-wave pairing in
silicene monolayers and bilayers, respectively. However, they have considered not phonon-mediated processes but, rather, an RPA multi-orbital Hubbard-model 
approach\cite{{Takimoto_2004},{Yada_2005},{Kubo_2007},{Mazin_2008},{Kuroki_2008},{Graser_2009}}. 
   
The {\it ab initio} calculations by Wan {\em et al.}\cite{Wan_2013} and Durajski and co-workers\cite{Durajski_2015} 
are for silicene under large tensile biaxial strain and result in an estimated $T_{\rm c}$ of about 10-to-20~K at densities exceeding 
$10^{14}$ cm$^{-2}$. Unfortunately, the sophistication afforded by first-principles calculations often comes at the price of a numerical
complexity that forces the use of additional approximations. For example, in Ref.~\onlinecite{Wan_2013} the very coarse mesh used to discretize the
Brillouin zone ($120 \times 120$) makes it impossible to capture correctly the long-wavelength behavior of the electron/ZA-phonon matrix elements, 
thus missing or underestimating {\em the} major physical effect we consider here. Moreover, the integration over the Fermi surface is performed by 
replacing the width of the shell with a Gaussian `smearing' with a width of 0.01~Ry, an energy that is much larger than any other energy of interest in the 
problem. As a result, Wan {\em et al.} find a transition temperature that increases with increasing density, despite the fact that they identify $\Gamma$-phonons 
(optical and ZA) as controlling the attractive effective electron-phonon interaction. 
In this case (very plausible and expected, given the divergence at long wavelengths of the electron/ZA-phonon interaction), 
one would expect the role of these $q=0$-phonons to grow as the radius of the Fermi circle shrinks; that is, at low densities. 
We should mention that Ezawa\cite{Ezawa_2012} has also speculated about the topological-superconductor nature of some
`popular' non-$\sigma_{\rm h}$-symmetric 2D crystals, such as silicene, germanene, and stanene. 
Baskaran\cite{Baskaran_2016} has similarly argued about possible room-temperature superconductivity of silicene and germanene. 
Given this state of affairs, it is worth revisiting the problem, following the same path that we have followed so far, attempting to capture the long-wavelength region as accurately as possible.

\subsection{The full gap-equation}
\label{subsec:gapeq}

It is convenient to recast the gap equation, Eq.~(\ref{eq:gap_BCS}) in terms of the `frequency' variables
$\omega = \upsilon_{\rm F} k$, $\omega' = \upsilon_{\rm F} k'$, $\omega_{q} = b \ q^{3/2}$.
Then, gap equation can be written as:
\begin{multline}
\hspace*{-0.35cm}
\Delta(\omega) = \ - \frac{1}{(4 \pi \upsilon_{\rm F})^{2}} \int_{0}^{\omega_{\rm max}} {\rm d} \omega' \ \omega' \
                    \frac{\Delta(\omega')}{\sqrt{\hbar^{2}(\omega'-\omega_{\rm F})^{2} + \Delta(\omega')^{2}}} \\ 
                      \times {\mathcal{P}} \int_{\omega_{q,\rm min}}^{\omega_{q,{\rm max}}} {\rm d} \omega_{q} \ (1+\cos \phi ) \
                           \left ( \frac{{\rm d} \phi}{{\rm d} \omega_{q}} \right ) \ V(\omega,\omega',\omega_{q}) \ ,
\label{eq:gapNV}                    
\end{multline}
where the interaction potential is:
\begin{multline}
\hspace*{-0.35cm}
V(\omega,\omega',\omega_{q}) =
    - \frac{(DK_{0})^{2}}{\rho} \frac{\sin^{2}(\phi/2)}{\omega_{q}^{2}-(\omega'-\omega)^{2}} 
             \frac{\omega_{q}^{4/3}}{( \omega_{q}^{2/3}+\omega_{\beta}^{2/3} )^{2}} \\  
                  + \frac{e^{2}b^{2/3}}{\epsilon_{s}(\omega_{q}^{2/3}+\omega_{\beta}^{2/3})} \ .
\label{eq:VNV}
\end{multline} 
In Eq.~(\ref{eq:gapNV}), ${\mathcal{P}}$ denotes the Cauchy principal part of the integral, and:
\begin{equation}
\left \{
\begin{array}{ll}
\omega_{\rm max} & = 2 \pi \upsilon_{\rm F}/a_{0} \\ 
&\\
\omega_{q,{\rm max/min}} & =  \dfrac{b}{\upsilon_{\rm F}^{3/2}} \ |\omega' \pm \omega|^{3/2} \\
&\\
\omega_{\beta} & = b \ \beta^{3/2} \\
&\\
\cos \phi  & = \dfrac{\omega^{2}+\omega'^{2}-(\upsilon_{\rm F}^{2}/b^{4/3}) \ \omega_{q}^{4/3}}{2 \ \omega \ \omega'} \\
&\\
\dfrac{{\rm d}\phi}{{\rm d}\omega_{q}} 
   & = \dfrac{4}{3} \dfrac{\upsilon_{\rm F}^{2}}{b^{4/3}} \dfrac{\omega_{q}^{1/3}} {2 \ \omega \ \omega' \sin \phi} \ .
\end{array}
\right. 
\label{eq:variables}
\end{equation}
Of course, $\phi$ is expressed as a function of $\omega$, $\omega'$ and $\omega_{q}$, and so are $\sin^{2}(\phi/2)= (1-\cos \phi)/2$ and
$\sin \phi = \sqrt{1-\cos^{2}(\phi)}$. 

Equation~(\ref{eq:gapNV}), as well as Eq.~(\ref{eq:gap_Matsubara_final}) below, 
is a slightly simplified form of the Eliashberg's equation (see, for example, Eq.~(37b) of Ref.~\onlinecite{Eliashberg_1960}): In addition to having
ingored the phonon Bose factors, consistently with the approximation embraced initially, and approximated the electron self-energy,
we have also ignored corrections to the phonon self-energy. This seems to be a common approximation in the Eliashberg's formalism, although, 
given the strength of the electron/ZA-phonon interaction in our case, these are corrections whose importance should be investigated. We shall ignore this 
issue here. Of course, in addition to these assumptions, the specific form of the electron-phonon interaction also differs, since Eliashberg considered 
metals with an electron-phonon matrix element growing linearly with $q$.
    
Obviously, we have not resolved the problem of the non-analyticity of the integrand function; we have simply transferred the essential singularities
from the denominator $\omega_{q}^{2} - (\omega' - \omega)^{2}$ in Eq.~(\ref{eq:VNV}) to essential singularities elsewhere, namely, the fractional
powers and absolute values in Eqs.~(\ref{eq:variables}). However, having formulated the gap equation in terms of these new variables,
we can identify the poles of the effective interaction potential, Eq.~(\ref{eq:VNV}), and attempt a numerical approach by  
considering Eq.~(\ref{eq:gapNV}) simplified, as usual, by assuming an $\omega$-independent gap and looking for the gap at
the Fermi surface, $\Delta(\omega_{\rm F})$:
\begin{multline}
\hspace*{-0.35cm}
1 = \ - \frac{1}{(4 \pi \upsilon_{\rm F})^{2}} \int_{0}^{\omega_{\rm max}} {\rm d} \omega' \ \omega' \
                    \frac{1}{\sqrt{\hbar^{2}(\omega'-\omega_{\rm F})^{2} + \Delta(\omega_{\rm F})^{2}}} \\ 
                      \times {\mathcal{P}} \int_{\omega_{q,\rm min}}^{\omega_{q,{\rm max}}} {\rm d} \omega_{q} \ (1+\cos \phi ) \
                           \left ( \frac{{\rm d} \phi}{{\rm d} \omega_{q}} \right ) \ V(\omega_{\rm F},\omega',\omega_{q}) \ ,
\label{eq:gapNV_F}                    
\end{multline}
considering also the equation for the transition temperature, $T_{\rm c}$:
\begin{multline}
\hspace*{-0.35cm}
1 = \ - \frac{1}{(4 \pi \upsilon_{\rm F})^{2}} \int_{0}^{\omega_{\rm max}} {\rm d} \omega' \ \omega' \
                    \frac{1}{\hbar|\omega'-\omega_{\rm F}|} \ \tanh \left ( \frac{\hbar |\omega'-\omega_{\rm F}|}{2 k_{\rm B} T_{\rm c}} \right ) \\ 
                    \times {\mathcal{P}} \int_{\omega_{q,\rm min}}^{\omega_{q,{\rm max}}} {\rm d} \omega_{q} \ (1+\cos \phi ) \
                           \left ( \frac{{\rm d} \phi}{{\rm d} \omega_{q}} \right ) \ V(\omega_{\rm F},\omega',\omega_{q}) \ .
\label{eq:TcNV_F}                    
\end{multline}
\begin{figure}[tb]
\includegraphics[width=8.0cm]{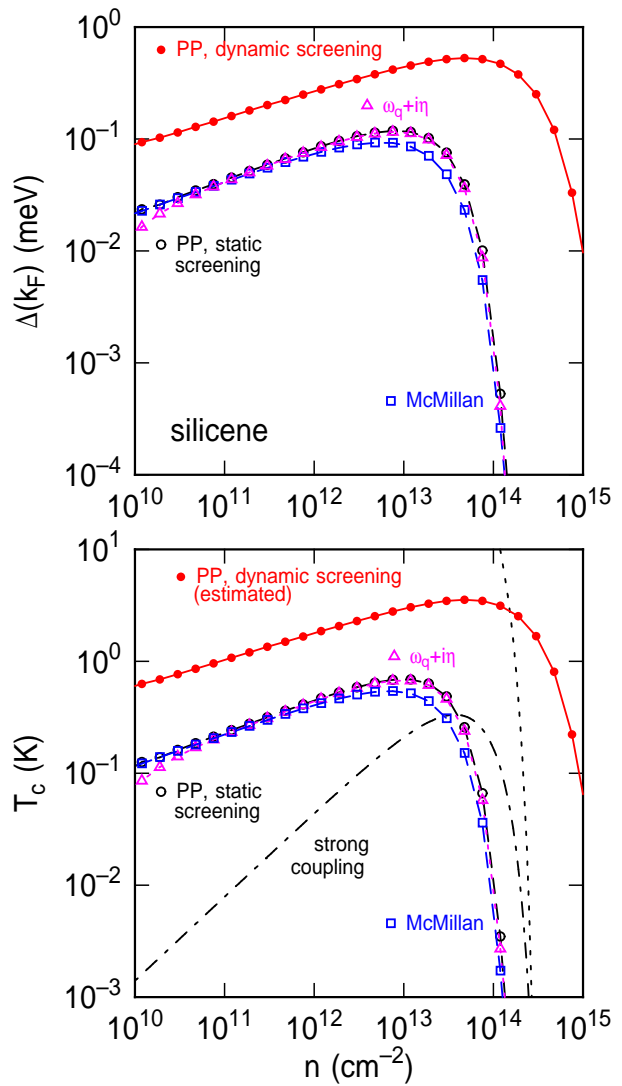}
\caption{Superconducting gap at the Fermi energy (top) and transition temperature (bottom) calculated for silicene using the various
approximations described in the text: The weak coupling limit, the strong-coupling limit, and the solution of the gap equation (constant-gap approximation)
integrating it through the singularities by computing the Cauchy principal part of the integral ('PP, static screening', black open circles); 
by regularizing the singularities accounting for a finite phonon lifetime, $1/\eta$, with $\eta = 10^{-6} \omega_{\rm D}$ 
(`$\omega + i \eta$', open triangles, cyan on-line); and 
employing McMillan's approximation, as described in the text (`McMillan', open squares, blue on-line). The dots (red on-line) also show the
results obtained accounting for dynamically screened interactions and calculating the Cauchy principal part of the integral.}
\label{fig:Cauchy_PP}
\end{figure}
\begin{figure}
\includegraphics[width=8.0cm]{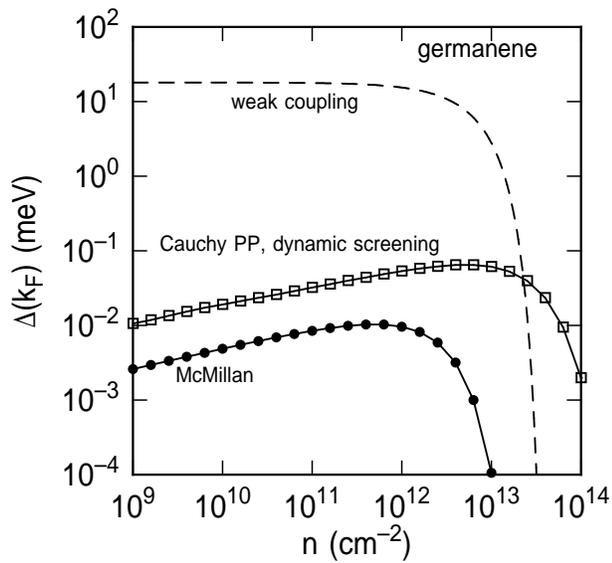}
\caption{Superconductivity gap for germanene calculated using some of the approximations described in the text and in Fig.~\ref{fig:Cauchy_PP}.}
\label{fig:germanene}
\end{figure}
Extreme care must be taken in order to treat correctly the Cauchy principal part of the integral, employing highly nonuniform meshes to discretize
the $\omega'$ and $\omega_{q}$ integration interval. 

Given the complexity of the numerical task at hand, we can confirm the correctness of our results by employing two additional strategies. The 
singularities can be regularized by adding an imaginary term to the phonon energy, $\omega_{q} \rightarrow \omega_{q} + i \eta$. Physically,
the lifetime $1/\eta$ may be thought as caused by the anharmonic coupling to in-plane phonons. The numerical integration can then be performed
in a similar way. Alternatively, we can adopt McMillan's approximation: In the denominator of Eq.~(\ref{eq:VNV}), we retain only the  
term $\omega_{q}^{2}$ when $\omega_{q}^{2} \ge (\omega'-\omega)^{2}$, or retain only the term $(\omega'-\omega)^{2}$ when
$\omega_{q}^{2} < (\omega'-\omega)^{2}$. 

Figure~\ref{fig:Cauchy_PP} shows the resulting gap (top frame) and transition temperature (bottom frame)
obtained by computing the Cauchy principal part of the $\omega_{q}$-integral in
Eq.~(\ref{eq:gapNV_F}) (open circles, black), as well as the two additional approximate solutions: The solution obtained by regularizing the
singularities by accounting for a finite phonon lifetime with $\eta = 10^{-6} \omega_{\rm D}$ (open triangles, cyan on-line) and using McMillan's
approximation (open squares, blue on-line). Note that the results obtained using these three different numerical strategies are in excellent agreement.
This is a non-trivial result, since it shows -- albeit indirectly -- that Migdal's theorem is likely to hold even in our rather unconventional case.
Figure~\ref{fig:germanene} shows the superconducting gap at the Fermi energy obtained for germanene using the various approximations we have considered
also for silicene. 

\subsection{The frequency-dependent gap}
\label{subsec:Deltaomega}
\begin{figure}[tb]
\includegraphics[width=8.0cm]{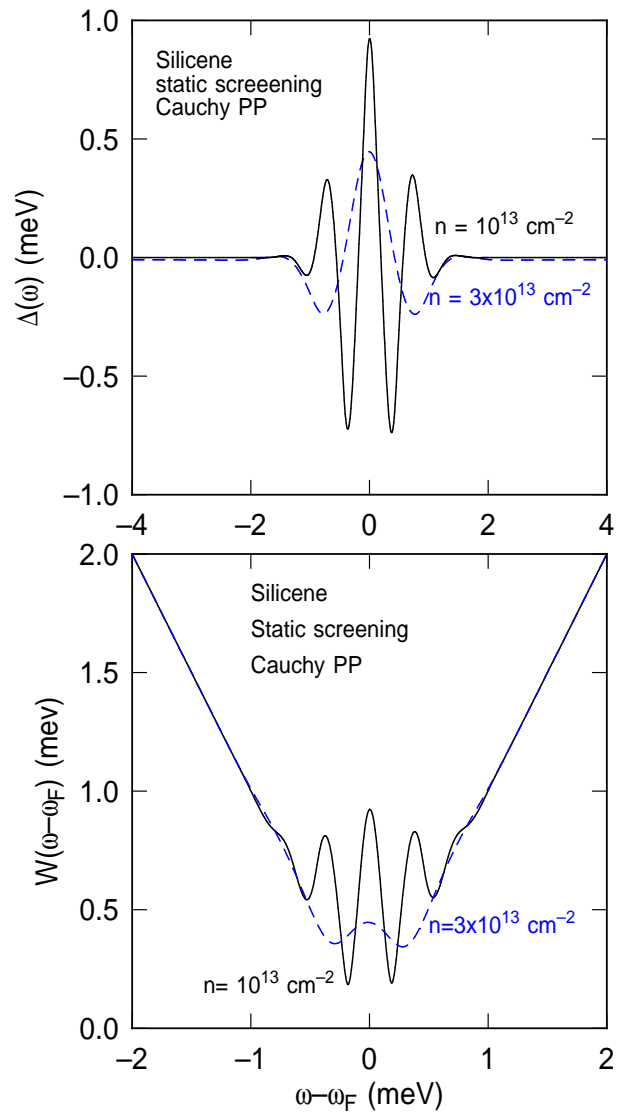}
\caption{Top: Superconducting gap as a function of frequency calculated by solving the integral gap equation at the indicated carrier densities. 
Bottom: Renormalized quasiparticle dispersion corresponding to the results shown in the top frame.}
\label{fig:dispersion}
\end{figure}
An assumption we have made consistently up to this point has been the `constant gap approximation'. Given the narrow range of ${\bf k}$-space in
which the effective electron-electron interaction is nonzero, we expect that this is a satisfactory approximation. More specifically, we expect
that the value of the gap (and transition temperature) obtained within this approximation will give some sort of average of the gap in the
neighborhood of the Fermi energy. Yet, it is interesting to confirm the correctness of our expectations by solving the full integral equation, Eq.~(\ref{eq:gapNV}). Given the significant computational cost, we have `spot-checked' our results
in the interesting range of carrier densities that yield the highest values for the gap in silicene. The integral equation is solved 
iteratively, starting from an initial guess of a Gaussian $\Delta(\omega)$ with peak value given by the result of the constant-gap
approximation and width given by the Debye frequency. Using the same integration mesh described above, convergence up to $10^{-6}$~eV is usually reached in
as few as 8-to-10 iterations. As shown in the top frame of Fig.~\ref{fig:dispersion}, at the indicated values 
the gap oscillates with a peak positive value at the Fermi
energy that is about an order of magnitude larger than the result of the constant-gap approximation. This latter result, as expected, provides an 
average value. The bottom frame of Fig.~\ref{fig:dispersion} illustrates the renormalized quasiparticle dispersion, showing that the minimum gap away from the Fermi energy is approximately of the same magnitude obtained using the constant-gap approximation.

\subsection{Dynamic screening}
\label{subsec:dynamic}
So far, we have considered interactions that are statically screened. As anticipated, we expect that dynamic screening may modify the picture in the
superconducting state. 

In order to account for dynamically screened interactions, we follow the standard procedure outlined, for example, in Ref.~\onlinecite{Sodemann_2012}.
We can re-express the equation for the gap $\Delta(\omega,i \omega_{n})$ in terms of a sum over Matsubara frequencies $\omega_{n}$ and screen the interaction potential using an analytic extension of the Wunsch' polarizability $\Pi^{\rm (RPA)}(q,\omega)$ to imaginary frequencies.
Restricting our attention to the gap calculated at $\omega_{n}=0$ and also assuming that it does not depend on $i \omega_{n}$, 
(an assumption that is just an extension of the `constant gap' approximation we have used before), the sum over the Matsubara frequencies can 
be converted to an integral over the imaginary axis in the zero-temperature limit. Rotating the integration axis to real frequencies and 
ignoring the imaginary part of $\Delta(\omega)$, we finally obtain:  
\begin{multline}
\hspace*{-0.35cm}
\Delta(\omega) = \ - \frac{1}{(4 \pi \upsilon_{\rm F})^{2}}   
             \int_{0}^{\omega_{\rm max}} {\rm d} \omega' 
                    \frac{\omega' \Delta(\omega')}{\sqrt{\hbar^{2}(\omega'-\omega_{\rm F})^{2} + \Delta(\omega')^{2}}} \\ 
                      \times {\mathcal{P}} \int_{\omega_{q,\rm min}}^{\omega_{q,{\rm max}}} {\rm d} \omega_{q} \ (1+\cos \phi ) \
                           \left ( \frac{{\rm d} \phi}{{\rm d} \omega_{q}} \right ) \\
                             \times \mbox{Re} \ V^{\rm (RPA)}(\omega,\omega',\omega_{q}; \widetilde{\omega}_{0}) \ .
\label{eq:gap_Matsubara_final}
\end{multline}
In this equation, note the presence of the gap $\Delta(\omega')$ in the frequency 
$\widetilde{\omega}_{0} = \sqrt{(\omega' - \omega_{\rm F})^{2} + \Delta(\omega')^{2}/\hbar^{2}}$ 
entering the RPA dielectric function. This had been anticipated in Sec.~\ref{subsec:nonBCS} in our early discussion about dielectric screening. 

Equation~(\ref{eq:gap_Matsubara_final}) evaluated at the Fermi energy, and also assuming the `usual' constant-gap approximation, becomes:
\begin{multline}
1 = \ - \frac{1}{(4 \pi \upsilon_{\rm F})^{2}}   
             \int_{0}^{\omega_{\rm max}} {\rm d} \omega' 
                    \frac{\omega'}{\sqrt{\hbar^{2}(\omega'-\omega_{\rm F})^{2} + \Delta(\omega_{\rm F})^{2}}} \\ 
                      \times {\mathcal{P}} \int_{\omega_{q,\rm min}}^{\omega_{q,{\rm max}}} {\rm d} \omega_{q} \ (1+\cos \phi ) \
                           \left ( \frac{{\rm d} \phi}{{\rm d} \omega_{q}} \right ) \\
                             \times \mbox{Re} \ V^{\rm (RPA)}(\omega_{\rm F},\omega',\omega_{q}; \widetilde{\omega}_{0}) \ ,
\label{eq:gap_Matsubara_constant_gap}
\end{multline}
where (to be explicit, just for completeness and clarity) $V^{\rm (RPA)}(\omega_{\rm F},\omega',\omega_{q}; \widetilde{\omega}_{0}) $ is given by 
Eq.~(\ref{eq:VNV}), but modified to account for dynamic screening:
\begin{multline}
V^{\rm (RPA)}(\omega_{\rm F},\omega',\omega_{q}; \widetilde{\omega}_{0}) = \\
    - \frac{(DK_{0})^{2}}{\rho} \frac{\sin^{2}(\phi/2)}{\omega_{q}^{2}-(\omega'-\omega_{\rm F})^{2}} 
             \frac{\omega_{q}^{4/3}}{[ \omega_{q}^{2/3}+\omega_{\beta}(\widetilde{\omega}_{0})^{2/3} ]^{2}} \\  
                 + \frac{e^{2}b^{2/3}}{\epsilon_{s}[\omega_{q}^{2/3}+\omega_{\beta}(\widetilde{\omega}_{0})^{2/3}]} \\ 
    = - \frac{(DK_{0})^{2}}{\rho} \frac{\sin^{2}(\phi/2)}{b^{2}q^{3}-(\omega'-\omega_{\rm F})^{2}} 
             \frac{q^{2}}{[q+\beta(\widetilde{\omega}_{0})]^{2}} \\  
                  + \frac{e^{2}}{\epsilon_{s}[q+\beta(\widetilde{\omega}_{0})]} \ ,
\label{eq:VNV_F_dynamic}
\end{multline} 
with $\beta(\widetilde{\omega}_{0}) = -e^{2}/(2 \epsilon_{0} q)  \ \Pi^{\rm (RPA)}(q,\widetilde{\omega}_{0})$, 
$\omega_{\beta}(\widetilde{\omega}_{0}) = b \ \beta(\widetilde{\omega}_{0})^{3/2}$ and $q = (\omega_{q}/b)^{2/3}$.
Unfortunately, a similar `simple' equation does not hold for the transition temperature, since we have taken the zero-temperature limit to
convert the sum over Matsubara frequencies to a numerically more convenient integral. 
Yet, the value for $\Delta(\omega_{\rm F})$ obtained by solving Eq.~(\ref{eq:gap_Matsubara_constant_gap}) can give us a qualitative idea of how dynamic screening 
may affect the normal-superconducting transition.

The results (symbols labeled 'PP, dynamic screening') are shown in Figs.~\ref{fig:Cauchy_PP} and \ref{fig:germanene} 
for silicene and germanene, respectively.
The effect is significant: The maximum values of the superconducting gap and transition temperature are shifted at larger carrier densities, but 
are also increased by almost one order of magnitude, the 'estimated' (using the BCS `universal' gap-to-temperature ratio)
critical temperature reaching a value of about 5~K for silicene. 

\subsection{Effect of the dielectric environment}
\label{subsec:Epsilon}
All results presented so far have been obtained in the ideal case of free-standing layers. It is interesting to have a qualitative idea of what effect
a substrate or gate dielectric may have on the superconducting gap. We know that the dielectric constant of 2D crystals tends to approach the value of
the dielectric constant of the supporting substrate and/or of the gate insulator (or passivating layer)\cite{Fang_2016}. This is the result of the fact 
that the polarization of the 2D layer is strongly affected by the polarization of the dielectric environment itself. The dielectric response of the 2D 
layer will also be different along the in-plane or out-of-plane directions\cite{Fang_2016}. Let's ignore these complications and, rather than tackling the
hard problem of calculating the full dielectric response of the system substrate/silicene/gate insulator, let's assume that the dielectric constant of
silicene takes a value similar to the average value of the system. Therefore, we can assume that silicene supported and passivated by hBN or by SiO$_{2}$
will have an isotropic dielectric constant $\epsilon_{\rm s} \approx$ 4 $\epsilon_{0}$ or $\approx$ 10 $\epsilon_{0}$, 
when supported by hBN/SiO$_{2}$ and gated by HfO$_{2}$.

\begin{figure}[tb]
\includegraphics[width=8.0cm]{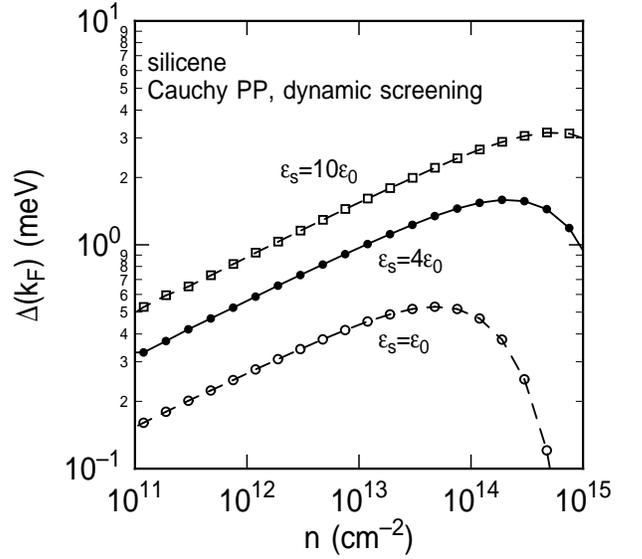}
\caption{Superconducting gap calculated by integrating the singularities via the Cauchy principal part and with a dynamic modle for screening assuming
different values for the silicene dielectric constant $\epsilon_{\rm s}$, resulting from different substrates or gate oxides.}
\label{fig:epsilon}
\end{figure}
 
Figure~\ref{fig:epsilon} shows the superconducting gap in these two over-simplified cases calculated using our `best' model; that is: integration of the
singularities by computing numerically the Cauchy principal part of the integral and employing dynamic screening. The more efficient screening of the 
Coulomb repulsive interaction and the reduced Thomas-Fermi screening length contribute to enhancing the gap. A  maximum value of about 2-3~meV can be
reached at the highest density for which our simplified band-structure remains valid ($\sim 7 \times 10^{13}$ cm$^{-2}$) and also reasonably achieved
in gated layers. Extrapolating from the `universal' ratio $k_{\rm B} T_{\rm c}/\Delta(k_{\rm F}) \sim 0.56$ seen in most of the cases we have considered, we can estimate that this will correspond to a transition temperature of about 15-20~K. Therefore, as long the the supporting substrate or gate insulator
couple with the 2D layer via weak Van der Waals forces that do not damp or suppress the flexural modes, a 'high-k' environment may be beneficial as far as
superconductivity is concerned.

\section{Conclusions}
\label{sec:Conclusions}
We have considered the strong interaction between electrons and flexural models that is allowed at first in non-$\sigma_{\rm h}$-symmetric
two-dimensional crystals and its potential to induce superconducting electron pairing. We have argued that in `gapped' materials that exhibit a parabolic 
electron dispersion, the interaction is too weak to lead to any superconductivity at realistic carrier densities. On the contrary, in 2D materials
that exhibit a Dirac-like electron dispersion, the interaction is strong enough to induce superconductivity. We have shown that this interaction is
significantly different from the interactions considered by the `conventional' BCS theory, since its strength increases at long wavelengths. 
Therefore, we have investigated the strong-coupling limit using McMillan's empirical strong-coupling (Eliashberg) formula, by solving directly the integral gap equation within the constant-gap approximation and assuming statically-screened interactions. We have shown that the negligible role played by
the singularities of the phonon-mediated electron-electron interaction suggests the validity of Migdal's theorem in our cases. Finally, we have
extended our study to account for dynamic screening, for the energy dependence of the superconducting gap, and for the presence of a dielectric environment
(supporting substrate or gate insulator). We have estimated that superconductivity in silicene can exhibit a transition temperature varying from 5~K to
20~K, depending on the dielectric environment. While not `earth-shaking', these results makes us conclude that the electron/ZA-phonon interaction should
be correctly accounted for when studying possible mechanisms leading to superconductivity in non-$\sigma_{\rm h}$-symmetric, Dirac-like 
two-dimensional materials. 
\vspace*{0.5cm}
\acknowledgments 
\vspace*{-0.5cm}
The authors would like to acknowledge constructive conversations with professor Alex Demkov and Mr. Donghan Shin (The University of Texas, Austin)
and with professor William Vandenberghe (The University of Texas, Dallas). 
This work has been partially supported by the Nanoelectronics Research Initiative's (NRI's) Southwest Academy of Nanoelectronics (SWAN) center affiliated with the Semiconductor Research Corporation.

\end{document}